\documentclass[sigconf]{acmart}

\usepackage[ruled,lined]{algorithm2e}
\usepackage{algorithmic}
\usepackage{subfig}
\usepackage{caption}
\usepackage{graphicx}
\usepackage{ragged2e} 
\usepackage{booktabs,makecell, multirow, tabularx}
\usepackage{threeparttable}
\usepackage{float}
\usepackage{caption}

\AtBeginDocument{%
  }

\setcopyright{acmcopyright}
\copyrightyear{2024}
\acmYear{2024}
\acmDOI{XXXXXXX.XXXXXXX}

\acmConference[Conference acronym 'XX]{Make sure to enter the correct
  conference title from your rights confirmation emai}{June 03--05,
  2024}{Woodstock, NY}
\acmPrice{15.00}
\acmISBN{978-1-4503-XXXX-X/18/06}

\begin{document}

\title{Node-based Knowledge Graph Contrastive Learning for Medical Relationship Prediction}

\author{Zhiguang Fan}
\affiliation{%
  \institution{Sun Yat-sen University}
  \city{GuangZhou}
  \country{China}}
\email{fanzhg5@mail2.sysu.edu.cn}

\author{Yuedong Yang}
\affiliation{%
	\institution{Sun Yat-sen University}
	\city{GuangZhou}
	\country{China}}
\email{yangyd25@mail.sysu.edu.cn}

\author{Mingyuan Xu$^*$}
\affiliation{%
	\institution{Guangzhou National Laboratory, China}
	\city{GuangZhou}
	\country{China}}
\email{mingyuan.xu.sci@gmail.com}

\author{Hongming Chen$^*$}
\affiliation{%
	\institution{Guangzhou National Laboratory, China}
	\city{GuangZhou}
	\country{China}}
\email{chen\_hongming@gzlab.ac.cn}

\renewcommand{\shortauthors}{Trovato et al.}

\begin{abstract}
  The embedding of Biomedical Knowledge Graphs (BKGs) generates robust representations, valuable for a variety of artificial intelligence applications, including predicting drug combinations and reasoning disease-drug relationships. Meanwhile, contrastive learning (CL) is widely employed to enhance the distinctiveness of these representations. However, constructing suitable contrastive pairs for CL, especially within Knowledge Graphs (KGs), has been challenging. In this paper, we proposed a novel node-based contrastive learning method for knowledge graph embedding, NC-KGE. NC-KGE enhances knowledge extraction in embeddings and speeds up training convergence by constructing appropriate contrastive node pairs on KGs. This scheme can be easily integrated with other knowledge graph embedding (KGE) methods. For downstream task such as biochemical relationship prediction, we have incorporated a relation-aware attention mechanism into NC-KGE, focusing on the semantic relationships and node interactions. Extensive experiments show that NC-KGE performs competitively with state-of-the-art models on public datasets like FB15k-237 and WN18RR. Particularly in biomedical relationship prediction tasks, NC-KGE outperforms all baselines on datasets such as PharmKG8k-28, DRKG17k-21, and BioKG72k-14, especially in predicting drug combination relationships. We release our code at https://github.com/zhi520/NC-KGE.
\end{abstract}



\maketitle

\section{Introduction}
Knowledge graphs (KGs) are intricate networks, representing the human knowledge, where nodes denote meaningful entities and edges signify their relationships. Specifically, Biomedical Knowledge Graphs (BKGs), such as TTD, PharmGKB, OMIM and DrugBank, start from the high-quality manually curated biomedicine knowledge and capture the intricate interactions among biomedical entities like genes, chemicals, and diseases. They play a pivotal role in many artificial intelligence applications including drug repositioning, adverse drug reaction analysis, and proteomics data analysis. Recently, Knowledge graph embedding methods were developed to generate compact, distributed representations of entities and relations for performance improvements. Once learned, these embeddings can be analyzed using diverse scoring techniques to provide probability scores for triplets, valuable for downstream tasks such as prediction, clustering, and visualization. Especially, biomedical relationship prediction, uncovers hidden associations between biological entities, enhancing disease understanding and biomarker discovery, illustrating biological pathways.

Recent KGE methods can be roughly categorized into structure-based embeddings and enhanced knowledge graph embeddings. The structure-based embeddings focus on the usage of knowledge graph facts and can be further categorized into 3 types. (1) translational distance models such as TransE \cite{bordes2013translating}, RotatE \cite{sun2019rotate}, QuatE \cite{zhang2019quaternion}, DualE \cite{cao2021dual} and HAKE \cite{zhang2020learning}. (2) semantic matching methods including RESCAL \cite{nickel2011three}, TATEC \cite{garcia2014effective}, DistMult \cite{yang2014embedding}, HolE \cite{nickel2016holographic}, ComplEx \cite{trouillon2016complex}, ANALOGY \cite{liu2017analogical} and SimplE \cite{kazemi2018simple}. (3) neural-based methods, especially GNN-based methods including RGCN \cite{schlichtkrull2018modeling}, SACN \cite{shang2019end}, KBGAT \cite{nathani2019learning}, A2N \cite{bansal2019a2n}, CompGCN \cite{vashishth2019composition} and SE-GNN \cite{li2022does}. The enhanced knowledge graph embeddings aim to use additional information for a better representation, including (1) text-enhanced knowledge embedding such as TEKE \cite{wang2016text}, DKRL \cite{xie2016representation}, KG-BERT \cite{yao2019kg}, (2) logic-enhanced embedding like KALE \cite{guo2016jointly}, (3) image-enhanced embedding and etc. Although these methods have gain improvements in powerful representation of entities in knowledge graphs, it is still challenging for biomedical relationship predictions due to knowledge bases from the biomedical domain are usually sparse, redundant and incomplete.

In the past, graph contrastive learning, a self-supervised method, has achieved significant success in generating generalized, transferable, and robust representations for graph structured data. This success has illuminated the path for learning knowledge graph embeddings. Essentially, contrastive learning aims to extract hidden information between samples by bringing similar samples closer and pushing dissimilar ones apart in latent space. Its core objective is to distinguish between a pair of representations from the two augmentations of the same sample (positives) apart from the k pairs of representation from the other (negative) samples. Constructing highly confident contrastive pairs is crucial for the discriminative power of contrastive learning models. However, due to the intricate structures within knowledge graphs, defining these contrastive pairs is challenging. Consequently, there are only limited attempts to integrate contrastive learning strategies with Knowledge Graph Embedding (KGE) methods.

SimKGC \cite{wang2022simkgc}, for instance, creates contrastive pairs using semantic similarity through language models, deviating from the previous graph contrastive learning models that fully mined information underlying graph structures. However, the effectiveness of contrastive KGE methods based on semantic similarity heavily relies on the specific language models used. 
Another approach, KGE-SymCL \cite{liang2023knowledge}, utilizes the semantic similarity of entities in relation-symmetrical positions to construct positive contrastive samples.
However, extracting the structure of symmetric relations is a tedious process and there is no significant improvement compared to SimKGC.
In biomedical knowledge graphs, entities represent gene codes, targets, or chemical compounds. The features generated by language models may lead to inaccurate semantic estimation in this context. Therefore, there is a pressing need for a more stable and universally applicable contrastive learning criterion tailored for biomedical knowledge graphs.

In our study, we introduced a new and versatile node-based contrastive method for knowledge graph embeddings, termed NC-KGE, specifically designed for predicting biomedical relationships. For a given fact triplet in knowledge graph, NC-KGE identifies triplets in the knowledge graph where one entity and the relation type remain the same as positive samples, while all other triplets are considered negative samples. By maximizing the similarity score between positive samples and minimizing it between negative ones using a modified classic contrastive learning loss, NC-KGE enhances the convergence speed during training and improve the performance of non-contrastive methods, such as CompGCN and SE-GNN. Additionally, we incorporated a relation-aware multi-head attention (RAMHA) mechanism into NC-KGE to enhance the utilization of relation semantics and interactions among relations and entities. We evaluated NC-KGE's performance in relation prediction on both general public datasets (FB15k-237 and WN18RR) and biomedical-focused knowledge graphs (PharmKG8k-28, DRKG17k-21, and BioKG72k-14). Our extensive experiments revealed that NC-KGE competes effectively with state-of-the-art models on general knowledge graphs, and surpasses all baselines on biomedical knowledge graphs, particularly excelling in predicting drug combination relationships.

\section{Related Work}
Knowledge Graph Embedding (KGE) aims to encode the entities and relations to the low dimensional vector or matrix space while maximally preserving its topological properties. Recent existing KGE models can be roughly categorized into structure based embedding and enhanced knowledge embeddings, reviewed by Singh et al. \cite{choudhary2021survey} and MINERVINI et al. \cite{palmonari2020knowledge}. In this work, structure-based knowledge graph embedding methods are more related than enhanced knowledge graph embeddings.

\subsection{Structure-based Knowledge graph embedding methods}
\textbf{Translation distance models} interpret relations as translations operation from a head node to an tailor node in latent space, e.g., TransE \cite{bordes2013translating}, TransH \cite{wang2014knowledge}, TransR \cite{lin2015learning} and etc. TransE was among the initial attempts to represent relations as addition operations between entities. TransH projects entities onto relation-specific hyperplanes, allowing entities to play different roles in various relations. TransR maps nodes and relations into distinct entities spaces and relation-specific spaces. RotaE \cite{sun2019rotate} treats the relation as a rotation operation. PairRE \cite{chao2020pairre} can encode complex relationships and multiple relationship patterns at the same time.  Moreover, HousE \cite{li2022house} involves a novel parameterization based on the designed Householder transformations for rotation and projection.

\textbf{Semantic matching models},  including RESCAL \cite{nickel2011three}, DistMult \cite{yang2014embedding}, ComplEx \cite{trouillon2016complex}, ConvE \cite{dettmers2018convolutional}, SimplE \cite{kazemi2018simple}, CrossE \cite{huang2022cross}, QuatE \cite{zhang2019quaternion}, DualE \cite{cao2021dual}, are developed based on similarity scoring functions. RESCAL utilizes a bilinear similarity function to compute the scores of knowledge triples and assumes that positive triples have higher scores than negative ones. DisMult simplifies the bilinear similarity function through using diagonal matrix. ComplEx further generalized DisMult by using complex embeddings and Hermitian dot products. Besides, the advantage of quaternion representations is leveraged by QuatE to enrich the correlation information between head and tail entities based on relational rotation quaternions. Inspired by it, DualE is proposed to gain a better expressive ability by projecting the embeddings in dual quaternion space.

\textbf{Neural-based methods}, including ConvE \cite{dettmers2018convolutional}, RGCN \cite{schlichtkrull2018modeling}, SACN \cite{shang2019end}, KBGAT \cite{nathani2019learning}, A2N \cite{bansal2019a2n}, CompGCN \cite{vashishth2019composition} and SE-GNN \cite{li2022does}. For example, ConvE introduces the use of convolutional layers to extract information. RGCN introduces a relation-specific transformation to integrate relation information with message aggregation. RGHAT \cite{DBLP:conf/aaai/ZhangZZ0XH20} incorporates a two-level attention mechanism, addressing relations and entities separately. KE-GCN \cite{yu2021knowledge} introduces a joint propagation method to update node and edge embeddings simultaneously. CompGCN proposes various composition operations for neighbor aggregation to model the structure pattern of multi-relational graph. RAGAT \cite{liu2021ragat} constructs separate message functions for different relations, which aims at exploiting the heterogeneous characteristics of knowledge graphs. SE-GNN \cite{li2022does}, with its three levels of semantic evidence, achieves in-depth knowledge representation by meticulously merging these layers through multi-layer aggregation, leading to highly extrapolative knowledge representations.

\subsection{Contrastive Learning on Knowledge Graph}
Graph Contrastive learning (GCL) operates by mining the hidden information in intra-data in a self-supervised manner. These methods, such as GRACE \cite{zhu2020deep}, GraphCL \cite{you2020graph}, AutoGCL \cite{yin2022autogcl} and iGCL \cite{liang2023graph} etc, has been proven highly successful in both nodes representing learning, relation prediction, classification, graph generation and anomaly detections. Recently, only a few researchers attempt to extend graph contrastive learning into knowledge graph embedding learning. SimKGC \cite{wang2022simkgc} tend to combine samples with high semantic similarity as positive pairs. This method heavily rely on semantic similarity predicted by the specific language models.
However, the language model may estimate sematic similarity inaccurately for the biomedical entities including gene codes, targets, or chemical compounds, making these methods fails for biomedical knowledge graphs. KGE-SymCL \cite{liang2023knowledge} utilizes the semantic similarity of entities in relation-symmetrical positions to construct positive contrastive samples. 
However, extracting the structure of symmetric relations is a tedious process and there is no significant improvement compared to SimKGC.

\section{Preliminary}
Knowledge Graph (KG) is composed of the fact triplets, denoted as $\mathcal{G}=\left\{\left(e_h, r, e_t\right) \mid e_h, e_t \in \mathcal{E}, r \in \mathcal{R}\right\}$, where $\mathcal{E}$ is the set of entities (i.e., nodes), $\mathcal{R}$ is the set of relations (i.e., edge types), $e_h, e_t$ represents the head and tail entity, respectively, and $r$ represents the relation between them. 
Relation Prediction, also known as Link Prediction or Knowledge Completion, involves predicting missing links or relationships in a knowledge graph. For a given head entity $e_h \in \mathcal{E}$ and a relation $r \in \mathcal{R}$, the objective is to identify the most suitable tail entity $e_t \in \mathcal{E}$, forming a new plausible triple $\left(e_h, r, e_t\right)$ within $\mathcal{G}$. Here, we approach this task by scoring all the candidates $\left\{\left(e_h, r, e_t^{\prime}\right) \mid e_t^{\prime} \in \mathcal{E}\right\}$, maximizing the scores for genuine triples and minimizing the scores for all the other candidates.

\section{NC-KGE Methods}
\begin{figure*}[htbp]
	\centerline{\includegraphics[width=7in]{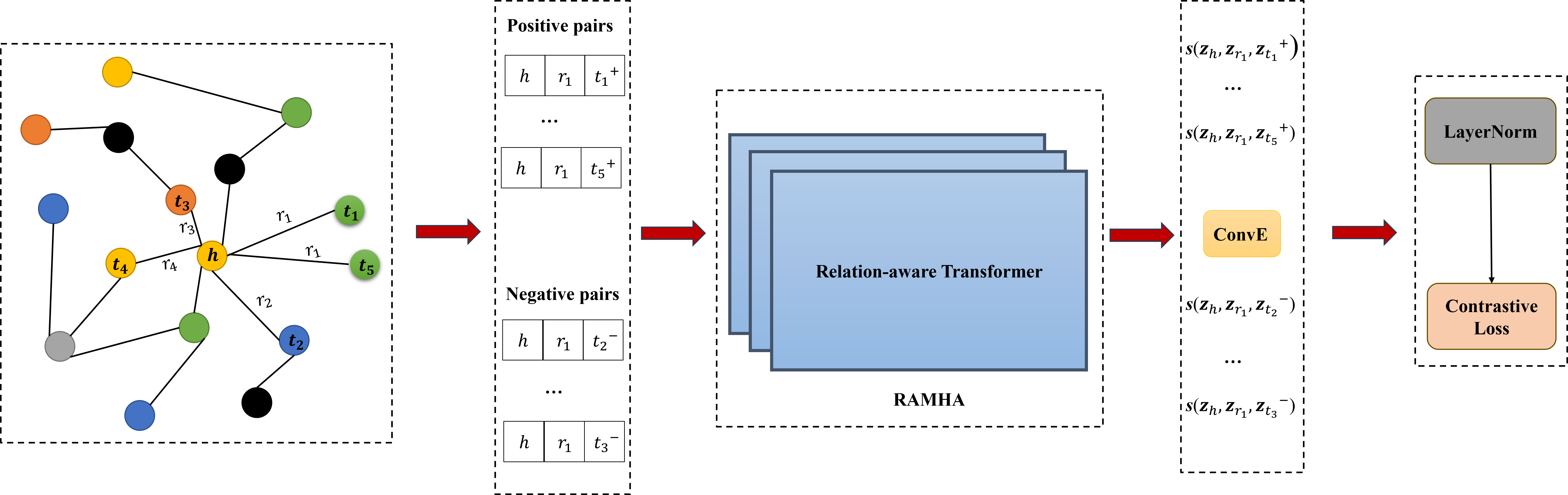}}
	\caption{The overall framework of NC-KGE.}
	\label{fig2}
\end{figure*}

As shown in Figure \ref{fig2}, we show the overall process of NC-KGE. For a given fact triple $\left(e_h, r, e_t\right)$ in biomedical knowledge graph $\mathcal{G}$, NC-KGE first constructs positive and negative augmented samples for contrastive learning. Then, a learnable KGE model was used to generate the embeddings for entities and relations in $\mathcal{G}$. Thirdly, a similarity function $S$ is used to score the triple embeddings $S\left(z_h, x_r, z_t^*\right)$ of both positive and negative samples, and calculates the contrastive loss to optimize KGE model by maximizing the scores of genuine triples and minimizing the scores for all the other candidates. During inference, NC-KGE measures the scores of triple embeddings $S\left(z_h, x_r, z_t^*\right)$ for all candidates $e_t^*$ when provided $\left(e_h, r, ?\right)$, and the candidate $e_t$ with highest scores will form a plausible triple with $e_h$ and $r$ in $\mathcal{G}$. Additionally, finding the head entity $e_h$ for the provided $\left(?, r, e_t\right)$ can be effortlessly transformed into a similar process.

\subsection{Construction of node-based contrastive pairs}
In biomedical knowledge graph $\mathcal{G}$, the entity $e_h$ can form the same relationship with multiple other entities $\left\{e_t \mid e_t \in \mathcal{E}\right\}$. For example, a target entity can correspond to multiple inhibitors, and one disease may be associated with multiple genes. For a given triple $\left(e_h, r, e_t\right)$, supporting another entity $e_t^{+}$ forms the same relation $r$ with the head entity $e_h$, $e_t^{+}$ is defined as a positive entity for $e_h$ and $r_i,\left(e_h, r, e_t^{+}\right)$ is a positive triple pair for $\left(e_h, r, e_t\right)$. Additionally, the triple $\left(e_h, r, e_t\right)$  is also regarded as a positive pair for itself. All the other entities $\left\{e_t^{-}\right\}$  which don’t form the relation $r$ with $e_h$ are defined as negative entity of $e_h$, and $\left(e_h, r, e_t^{-}\right)$ is also regarded as a negative pair for $\left(e_h, r, e_t\right)$.

\begin{figure*}[htbp]
	\centerline{\includegraphics[width=7in]{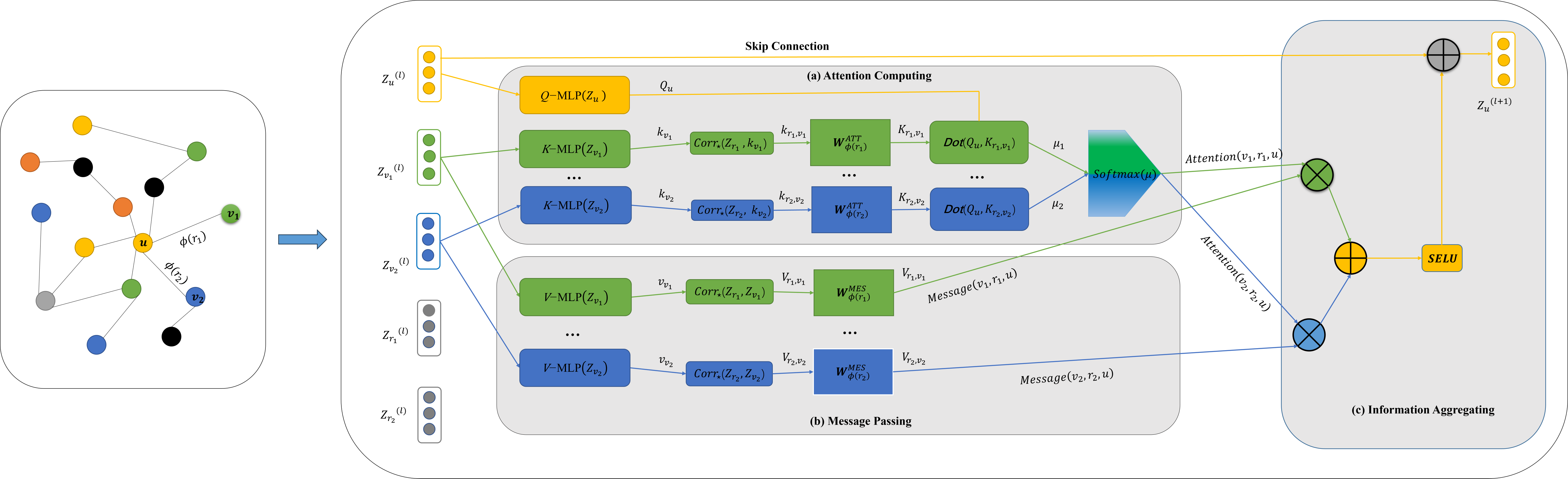}}
	\caption{Relation-aware multi-head attention based KGE model.}
	\label{fig3}
\end{figure*}

\subsection{Relation-aware multi-head attention based KGE model}
Here, we proposed a relation-aware multi-head attention mechanism (RAMHA) by integrating the relations between entities into attention computations to enhance the utilization of relation semantics and interactions among relations and entities in NC-KGE. As shown in Figure \ref{fig3}, RAMHA mainly includes three phases: attention computing, message passing and information aggregating. By stacking multiple layers of RAMHA, NC-KGE generates the knowledge graph embeddings for both entities and relations. 

Supporting the embedding of entity $e_u$ is denoted as $z_u^l$ and $x_r^l$  represents the embedding of relation type $r$ between $e_u$ and $e_v$ in $l$th layer of GAN, the next layer will aggregate information to $e_u$ from its neighbor $e_v$ according to their relations $r$ in an attention-based message passing process as shown in Equation \ref{eq1}. 
\begin{equation}
	z_u^{l+1}=\Bigg\|_{c=1}^C\Bigg(\sum_{v \in \mathcal{N}(u)} \sum_{r \in \mathcal{R}(u, v)} \alpha_{u, r, v}^{l+1, c} * Message_{u, r, v}^{l+1, c}\Bigg)
	\label{eq1}
\end{equation}
where $\alpha_{u, r, v}^{l+1, c}$ is the attention weight for the embedding triple $\left(z_u^l, x_r^l, z_v^l\right)$ for $c$th head in multi-head attentions, $Message_{u, r, v}$ denotes the information aggregated from $e_v$ to $e_u$ with relation $r$, $\|$ represents the concatenation operation between the total $C$ heads attentions. The embedding of relation r is updated in $l+1$ layer with Equation \ref{eq2} and \ref{eq3}.
\begin{equation}
	x_r^{l+1}=\Big\|_{c=1}^C(x_r^{l+1, c})
	\label{eq2}
\end{equation}

\begin{equation}
	x_r^{l+1, c}=M L P_r^{l, c}\left(x_r^l\right)
	\label{eq3}
\end{equation}
where $M L P_r^{l, c}$ is a relation-type specific MLP network in $l$th layer for $c$th heads. 

The attention weight $\alpha_{u, r, v}^{l, c}$ is as shown in Equation \ref{eq4} $\sim$ \ref{eq7}
\begin{equation}
	\alpha_{u, r, v}^{l+1, c}=\frac{\left\langle q_u^{l, c}, M L P_{r, 1}^{l, c}\left(k_v^{l, c} \star x_r^{l+1}\right)\right\rangle}{\sum_{w \in \mathcal{N}(u)} \sum_{r^{\prime} \in \mathcal{R}(u, w)}\left\langle q_u^{l, c}, M L P_{r^{\prime}, 1}^{l, c}\left(k_w^{l, c} \star x_{r^{\prime}}^{l+1, c}\right)\right\rangle}
	\label{eq4}
\end{equation}

\begin{equation}
	q_u^{l, c}=M L P_q^{l, c}\left(z_u^l\right)
	\label{eq5}
\end{equation}

\begin{equation}
	k_u^{l, c}=M L P_k^{l, c}\left(z_u^l\right)
	\label{eq6}
\end{equation}

\begin{equation}
	v_u^{l, c}=M L P_v^{l, c}\left(z_u^l\right)
	\label{eq7}
\end{equation}
where $q^{l, c}$ and $k^{l, c}$ is the query vector and key vector for entities in multi-head attentions, $M L P_{r, 1}^{l, c}$ is a relation-type specific MLP without any bias terms, $\mathcal{N}(u)$ represents the set of neighbors of entity $e_u$, $\mathcal{R}(u, w)$ denotes the set of relations between $e_u$  and $e_w$, $\langle q, k\rangle=\exp \left(\frac{q^{T_{* k}}}{\sqrt{d}}\right)$, and $\star$ represents the circular-correlation mentioned in HolE \cite{nickel2016holographic}.  

The message passing information Message $e_{u, r, v}^{l, c}$ is as follows:
\begin{equation}
	{Message}_{u, r, v}^{l+1, c}=M L P_{r, 2}^{l, c}\left(v_t^{l, c} \star x_r^{l+1, c}\right)
	\label{eq8}
\end{equation}
where $M L P_{r, 2}^{l, c}$  is also an unbiased relation-type specific MLP as $M L P_{r, 1}^{l, c}$.

\subsection{Similarity scoring functions for relation predictions}
Once embeddings of entities and relations are obtained, NC-KGE introduced a learnable similarity function to score the embeddings of triplets in the knowledge graph as same as ConvE. It is a convolution over 2D shaped embeddings as formulated in Equation \ref{eq9}.
\begin{equation}
	\psi\left(e_u, r, e_v\right)=f\left(\text { reshape }\left(f\left(\left[\overline{e_u}, \bar{r}\right] \otimes w\right)\right) * W\right) \cdot e_v
	\label{eq9}
\end{equation}
Where $\overline{e_u}, \bar{r}$ denote a 2D reshaping of $e_u$ and relation type $r$, [] denotes a concatenation operation, $\otimes w$ represents a 2D-convolutional layer with filters $w$, $f$ denotes a non-linear function, $W$ is a linear transformation matrix, and $\cdot$ represents the inner product operation. We need to note that NC-KGE can also integrate the classic statistic similarity measure functions as TransE, DisMult, ComplEx and SimplE as listed Table \ref{tab1}, but extensive experiments have shown that a learnable similarity function in Equation \ref{eq9} outperforms these traditional similarity measures, as discussed in Section 5.4 Results.

\begin{table}[htbp]
	\centering
	\caption{{Examples of triple similarity measure functions proposed by TransE, DistMult, ComplEx, SimplE and ConvE. For triples $\left(e_{i}, r_{j}, e_{k}\right)$, we use $\mathrm{e}_{i}, \mathrm{r}_{j}, \mathrm{e}_{k}$ to represent the embedding of its components (in SimplE, they have two parts, which we use index to represent). $\|\cdot\|_{p}$ represents the $p$-norm; $\langle\cdot, ;\rangle$ is the generalized three-way dot product; $\operatorname{Re}(\cdot)$ is the real part of the complex number; $\bar{e}_{k}$ is the complex conjugate of the complex-valued vector $\mathrm{e}_{k}$.}}
	\begin{tabular}{lc}
		\toprule
		Model & Function \\
		\midrule
		TransE \cite{bordes2013translating}&   $-\left\|\mathrm{e}_{i}+\mathrm{r}_{j}-\mathrm{e}_{k}\right\|_{p}$ \\
		DistMult \cite{yang2014embedding} &  $\left\langle\mathrm{e}_{i}, \mathrm{r}_{j}, \mathrm{e}_{k}\right\rangle$ \\
		ComplEx \cite{trouillon2016complex}&  $\operatorname{Re}\left(\left\langle\mathrm{e}_{i}, \mathrm{r}_{j}, \overline{\mathrm{e}}_{k}\right\rangle\right)$ \\
		SimplE \cite{kazemi2018simple} & $\frac{1}{2}\left(\left\langle\mathrm{e}_{i 1}, \mathrm{r}_{j 1}, \mathrm{e}_{k 1}\right\rangle+\left\langle\mathrm{e}_{i 2}, \mathrm{r}_{j 2}, \mathrm{e}_{k 2}\right\rangle\right)$  \\
		ConvE \cite{dettmers2018convolutional} & $f\left(vec\left(f\left(concat\left(\overline{e_i}, \overline{r_j}\right) * \omega\right)\right) W\right) e_k$ \\
		\bottomrule
	\end{tabular}%
	\label{tab1}%
\end{table}%

\subsection{Contrastive learning objective}
NC-KGE aims to mining the hidden information between entities and relations by maximizing the similarity score between positive samples and minimizing it between negative ones. Thus, a classic contrastive training objective function are introduced into NC-KGE for a given triplet $\left(e_h, r, e_t\right)$ as shown in Equation \ref{eq10}.
\begin{equation}
	\small
	\mathcal{L}=-\log \frac{\sum_{k=1}^{K^{+}} \exp \left(\bar{S}\left(z_h, x_r, z_t^{+}\right) / \tau\right)}{\sum_{k=1}^{K^{+}} \exp \left(\bar{S}\left(z_h, x_r, z_t^{+}\right) / \tau\right)+Q \sum_{k=1}^{K^{-}} \exp \left(\bar{S}\left(z_h, x_r, z_t^{-}\right) / \tau\right)}
	\label{eq10}
\end{equation}
where the construction $K^{+}$and $K^{-}$ are the number of positive triple pairs and negative triple pairs, respectively, $Q$ is a scaling weight for negative pairs, $\tau$ is a temperature factor controlling the discriminable capability of KGE model to negative pairs. To avoid extreme temperature coefficients affecting contrastive learning effects, we adopt a simulated annealing strategy to adjust the temperature factor dynamically in the range of [0.1,1.5] according to the MRR metrics. Additionally, we need to note that the similarity scores $S$ were layer-normalized to compute the contrastive loss, denoted as $\bar{S}$, to avoid the numerical overflow in exponential operation and underflow in logarithmic operations.

\section{Experiments}
Two commonly used benchmark datasets FB15k-237 and WN18RR \cite{daza2021inductive} for KGE methods are utilized to evaluate the performance of NC-KGE on relation-predictions. We also perform benchmarks on three Bio-medical specific datasets including PharmKG8k-28, DRKG17k-21 and BioKG72k-14, derived from PharmKG, Drug Repositioning Knowledge Graph, and BioKG datasets. In these benchmarks, we could further evaluate the performance of NC-KGE on relation-predictions between different bio-meaningful entities. The detailed description of datasets is shown in Table \ref{tab2} of Appendix A.

\subsection{Baselines}
The baselines are composed of three types: Translation Model, Semantic Matching Model and GNN-based Model. Translation Models include TransE \cite{bordes2013translating}, RotatE \cite{sun2019rotate} and PaiRE \cite{chao2020pairre}. 
Semantic Matching Models include DistMult \cite{yang2014embedding}, ComplEx \cite{trouillon2016complex}, TuckER \cite{balavzevic2019tucker}, ConvE \cite{dettmers2018convolutional}, InteractE \cite{vashishth2020interacte} and PROCRUSTES \cite{peng2021highly}. 
GNN-based Models include  HyConvE \cite{wang2023hyconve},  MEKER \cite{chekalina2022meker}, RAGAT \cite{liu2021ragat}, HRGAT \cite{10.1093/bib/bbaa344}, R-GCN \cite{schlichtkrull2018modeling}, KBGAT \cite{nathani2019learning}, A2N \cite{bansal2019a2n}, SACN \cite{shang2019end}, CompGCN \cite{vashishth2019composition} and SE-GNN \cite{li2022does}

\subsection{Task and Evaluation}
Relation Prediction, also termed Link Prediction, aims at inferring missing facts based on the facts in Knowledge Graph. Similar to question answering, we assess the quality of relation-predictions using the following ranking task: for all triplets $\left(e_h, r, e_t\right)$ in both training sets and test sets, (1) we hidden the tail entity $e_t$. (2) we compute the similarity scores $S\left(e_h, r, e_t^*\right)$ for all $e_t^* \in \mathcal{E}$ as discussed in Section 4.3. (3) we sort values by decreasing order and (4) record the rank of the correct entity $e_t$. An identical process is repeated for prediction $e_h$.

Two kinds of metrics are introduced for evaluation, including the proportion of correct entities ranked in the top (such as top 1,3,10, denoted as Hits@1, Hits@3, Hits@10), and the mean reciprocal rank (MRR). Let $r_{t h}$ be the rank of the correct triplet $t=\left(e_h, r, e_t\right)$ among all possible triples when hidden the head entity, while $r_{t t}$ represent its rank when hidden the tail entity. MRR is the average of reciprocal rank of a set of correct fact triplets $\mathcal{T}$ as shown in Equation \ref{eq11}.
\begin{equation}
	M R R=\frac{1}{2|\mathcal{T}|} \sum_{t \in \mathcal{T}} \frac{1}{r_{t h}}+\frac{1}{r_{t t}}
	\label{eq11}
\end{equation}
The Hits at 1 metric (H@1) is obtained by counting the number of times the correct triple appears at position 1. The H@3 and H@10 are computed similarly, considering
the first 3 and 10 positions, respectively.

\subsection{Experimental Setup}
In our benchmark experiments, we employed a KGE encoder in NC-KGE comprising two layers of RAMHA. The RAMHA model utilized 10 heads, and the hidden layer dimension for all MLP was 200. For Equation \ref{eq10}, the positive pair number for each fact triplet $K^{+}$ was set to 1 while all the negative pairs are used in node-based contrastive training. To stabilize the training phase and prevent overfitting, we introduced a dropout rate of 0.2 and applied batch layer normalization between the RAMHA layers. The AdamW \cite{loshchilov2017decoupled} optimizer and cosine annealing learning rate scheduler \cite{loshchilov2016sgdr} is used in training. The patience of simulated annealing strategy for temperature factor is set to 50 according to MRR metric.

\begin{table*}[htbp]
	\centering
	\caption{Model reports on FB15k-237 and WN18RR test set. The best results are in bold. $^\dagger$ denotes that results are from the published paper. Other results are from SE-GNN \cite{li2022does}.}
	\begin{tabular}{lcccc|cccc}
		\toprule
		\textbf{Dataset} & \multicolumn{4}{c}{\textbf{FB15k-237}} & \multicolumn{4}{c}{\textbf{WN18RR}} \\
		\midrule
		\textbf{Task} & \multicolumn{4}{c}{\textbf{Link Prediction}} & \multicolumn{4}{c}{\textbf{Link Prediction}} \\
		\midrule
		\textbf{Metric $\rightarrow$} & MRR   & H@1  & H@3    & H@10   & MRR   & H@1   & H@3   & H@10 \\
		\textbf{Model $\downarrow$} &       &       &       &       &       &       &       &  \\
		\midrule
		TransE & 0.294  &  -    &  -    & 0.465  & 0.226  &  -    &  -    & 0.501  \\
		RotatE & 0.338  & 0.241  & 0.375  & 0.533  & 0.476  & 0.428  & 0.492  & 0.571  \\
		PaiRE & 0.351  & 0.256  & 0.387  & 0.544  &  -    &  -    &  -    &  - \\
		\midrule
		DistMult & 0.241  & 0.155  & 0.263  & 0.419  & 0.430  & 0.390  & 0.440  & 0.490  \\
		ComplEx & 0.247  & 0.158  & 0.275  & 0.428  & 0.440  & 0.410  & 0.460  & 0.510  \\
		TuckER & 0.358  & 0.266  & 0.394  & 0.544  & 0.470  & 0.443  & 0.482  & 0.526  \\
		ConvE & 0.325  & 0.237  & 0.356  & 0.501  & 0.430  & 0.400  & 0.440  & 0.520  \\
		InteractE & 0.354  & 0.263  &  -    & 0.535  & 0.463  & 0.430  &  -    & 0.528  \\
		PROCRUSTES & 0.345  & 0.249  & 0.379  & 0.541  & 0.474  & 0.421  & 0.502  & 0.569  \\
		\midrule
		HyConvE$^\dagger$ & 0.339  & 0.212  &  -    & 0.458  & 0.461  & 0.432  &  -    & 0.534  \\
		MEKER$^\dagger$ & 0.359  & 0.268  & 0.392  & 0.539  & 0.477  & 0.437  & 0.488  & 0.545  \\
		R-GCN & 0.248  & 0.151  &  -    & 0.417  &  -    &  -    &  -    &  - \\
		KBGAT & 0.157  &  -    &  -    & 0.331  & 0.412  &  -    &  -    & 0.554  \\
		A2N & 0.317  & 0.232  & 0.348  & 0.486  & 0.450  & 0.420  & 0.460  & 0.510  \\
		SACN & 0.350  & 0.260  & 0.390  & 0.540  & 0.470  & 0.430  & 0.480  & 0.540  \\
		CompGCN & 0.355  & 0.264  & 0.390  & 0.535  & 0.479  & 0.443  & 0.494  & 0.546  \\
		SE-GNN & 0.365 & 0.271  & \textbf{0.399 } & \textbf{0.549 } & 0.484  & 0.446 & \textbf{0.509 } & \textbf{0.572 } \\
		\midrule
		\textbf{NC-KGE (our)} & \textbf{0.366}  & \textbf{0.273}  & 0.392  & 0.542  & \textbf{0.486}  & \textbf{0.447}  & 0.499  & 0.556  \\
		\bottomrule
	\end{tabular}%
	\label{tab3}%
\end{table*}%

\begin{table*}[htbp]
	\centering
	\caption{The results of models on DRKG17k-21, BioKG72k-14 and PharmKG8k-28 Datasets. $^\dagger$ denotes that we reproduce the results using the codes \protect\footnotemark[1]. For other results, we implemented the official codes.}
	\begin{tabular}{l|cccc|cccc|cccc}
		\toprule
		\multicolumn{1}{l}{\textbf{Dataset}} & \multicolumn{4}{c}{\textbf{DRKG17k-21}} & \multicolumn{4}{c}{\textbf{BioKG72k-14}} & \multicolumn{4}{c}{\textbf{PharmKG8k-28}} \\
		\midrule
		\multicolumn{1}{l}{\textbf{Task}} & \multicolumn{4}{c}{\textbf{Link Prediction}} & \multicolumn{4}{c}{\textbf{Link Prediction}} & \multicolumn{4}{c}{\textbf{Link Prediction}} \\
		\midrule
		\textbf{Metric $\rightarrow$} & MRR   & H@1   & H@3   & H@10  & MRR   & H@1   & H@3   & H@10  & MRR   & H@1   & H@3   & H@10 \\
		\textbf{Model $\downarrow$} &       &       &       &       &       &       &       &       &       &       &       &  \\
		\midrule
		TransE  & 0.321  & 0.035  & 0.558  & 0.744  & 0.116  & 0.026  & 0.149  & 0.276  & 0.116  & 0.038  & 0.127  & 0.269  \\
		DistMult & 0.240  & 0.175  & 0.254  & 0.371  & 0.045  & 0.021  & 0.041  & 0.083  & 0.218  & 0.152  & 0.237  & 0.335  \\
		ComplEx & 0.099  & 0.036  & 0.087  & 0.227  & 0.111  & 0.073  & 0.118  & 0.174  & 0.124  & 0.064  & 0.128  & 0.244  \\
		TruckER & 0.460  & 0.411  & 0.535  & 0.557  & 0.226  & 0.174  & 0.237  & 0.327  & 0.182  & 0.103  & 0.202  & 0.336  \\
		HRGAT & 0.540  & 0.483  & 0.582  & 0.720  & 0.103  & 0.061  & 0.105  & 0.185  & 0.134  & 0.063  & 0.144  & 0.271  \\
		SACN  & 0.487  & 0.393  & 0.534  & 0.665  & 0.179  & 0.118  & 0.192  & 0.299  & 0.156  & 0.085  & 0.170  & 0.296  \\
		CompGCN & 0.562  & 0.466  & 0.619  & 0.739  & 0.221  & 0.170  & 0.230  & 0.321  & 0.193  & 0.110  & 0.216  & 0.352  \\
		SE-GNN & 0.575  & 0.481  & 0.631  & 0.746  & 0.237  & 0.183  & 0.248  & 0.343  & 0.206  & 0.120  & 0.232  & 0.374  \\
		\midrule
		\textbf{NC-KGE (our)} & \textbf{0.590 } & \textbf{0.505 } & \textbf{0.637 } & \textbf{0.747 } & \textbf{0.240 } & \textbf{0.185 } & \textbf{0.256 } & \textbf{0.344 } & \textbf{0.228 } & \textbf{0.145 } & \textbf{0.252 } & \textbf{0.390 } \\
		\bottomrule
	\end{tabular}%
	\label{tab5}%

\end{table*}%
\footnotetext[1]{https://github.com/DeepGraphLearning/}

\subsection{Results}
Here, we first evaluate the performance of NC-KGE on general knowledge graph FB15k-237 and WN18RR, the benchmark results are as shown in Table\ref{tab3}. Compared with 17 baselines, NC-KGE outperforms all the translation distance-based and semantic matching-based models, and is competitive with the SOTA method, SE-GNN. NC-KGE obtains obvious improvement compared to CompGCN, a typical GNN-based model, indicating node-based contrastive learning and RAMHA sufficient for better knowledge graph embedding.

For bio-medical specific knowledge graphs, NC-KGE outperforms all the baselines as shown in Table \ref{tab5}. In PharmKG8k-28, the knowledge entity can be clearly categorized into four bio-medical meaningful types: gene, disease and chemical compounds, Thus, the relation prediction can be divided into 6 types according to the types of head and tail entity of triplet. Then, we further evaluate the relation prediction performance of NC-KGE on these subdomains. As shown in Table \ref{tab10}, NC-KGE give obviously better results for Chemical-Chemical and Disease-Disease relation predictions, such as drug combination relationship and disease complications, which MRR is 0.483 and 0.471, respectively. Both of these subdomains are belonging to “Interactions” category in PharmKG.  Similarly, NC-KGE also achieve the best results, MRR is 0.640 on Chemical-Chemical relation predictions, since no Chemical-Disease and Disease-Disease relations exists as shown in Table \ref{tab11}.

\begin{figure*}[htbp]  
	\centering    
	
	\subfloat[Multi-model Convergence.] 
	{
		\begin{minipage}[t]{0.5\textwidth}
			\centering          
			\includegraphics[width=1\textwidth]{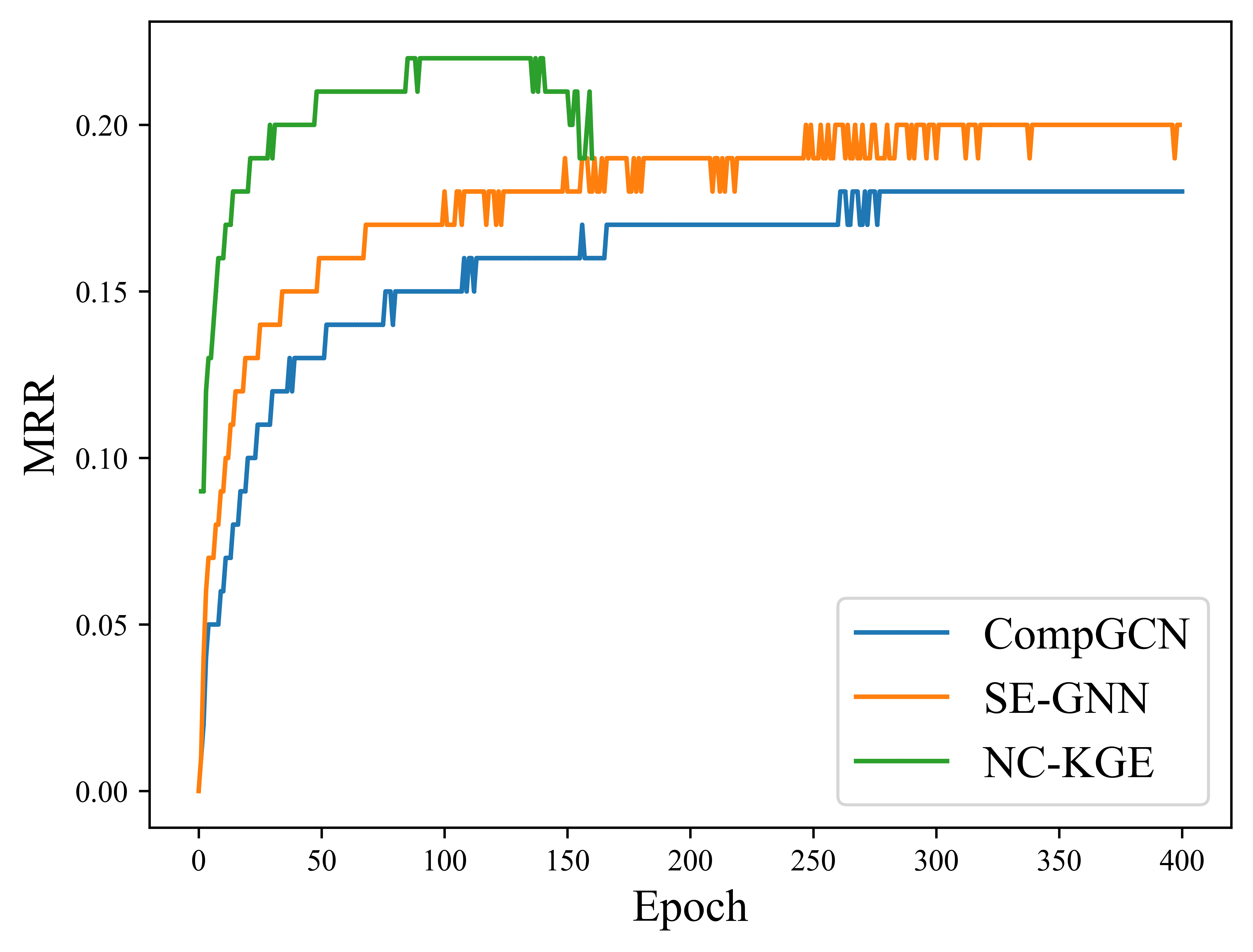}   
			\label{fig61}
		\end{minipage}%
	}
	\subfloat[Multi-loss Convergence on KGE.] 
	{
		\begin{minipage}[t]{0.5\textwidth}
			\centering          
			\includegraphics[width=1\textwidth]{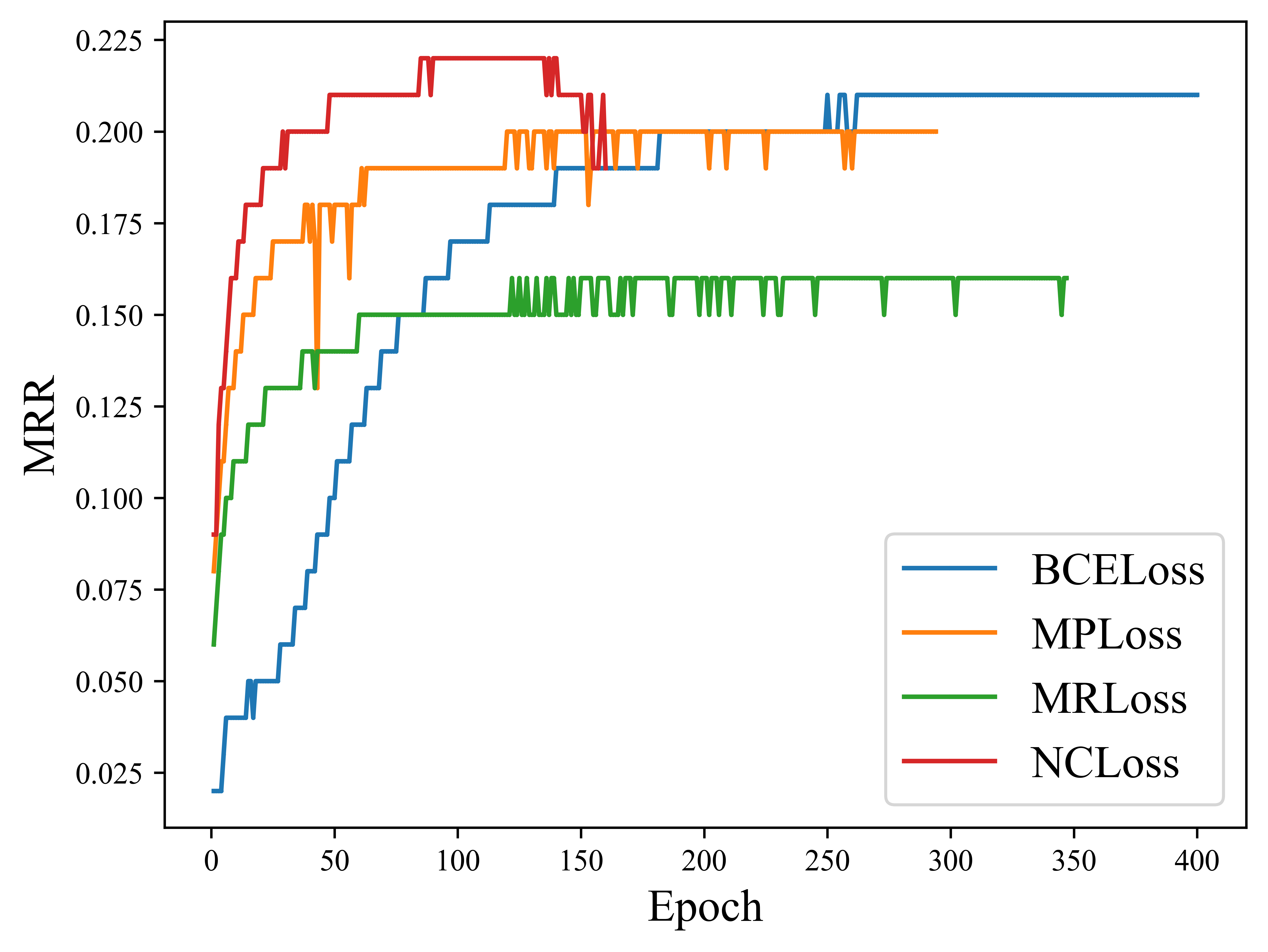}   
			\label{fig62}
		\end{minipage}%
	}
	\caption{Performance of the models on PharmKG8k-28.} 
	\label{fig6}  
\end{figure*}

Additionally, NC-KGE proves to be more efficient for training, as illustrated in Figure \ref{fig6},  Figure \ref{fig8}, and  Table \ref{tab8}. The Mean Reciprocal Rank (MRR) on test sets converges faster with NC-KGE than with any other methods. When we substitute the node-based contrastive objective with other functions like BCELoss (Binary CrossEntropy), MPLoss (Maximize Positive Sample Softmax Loss) and  MRLoss (Margin Ranking Loss), both accuracy and efficiency decrease. These findings highlight the efficiency and potential applications of NC-KGE in biomedical knowledge graphs.

\begin{table}[htbp]
	\centering
	\caption{Experiments on the PharmKG8k-28 subdataset}
	\begin{tabular}{lcccc}
		\toprule
		\textbf{Dataset} & \multicolumn{4}{c}{\textbf{PharmKG8k-28}} \\
		\midrule
		\textbf{Task} & \multicolumn{4}{c}{\textbf{Link Prediction}} \\
		\midrule
		\textbf{Metric $\rightarrow$} & MRR   & H@1   & H@3   & H@10 \\
		\textbf{SubDataset $\downarrow$} &       &       &       &  \\
		\midrule
		Gene-Chemical & 0.160  & 0.093  & 0.169  & 0.296  \\
		Chemical-Disease & 0.153  & 0.082  & 0.160  & 0.291  \\
		Disease-Disease & 0.471  & \textbf{0.377 } & 0.507  & 0.653  \\
		Gene-Gene & 0.244  & 0.156  & 0.275  & 0.413  \\
		Gene-Disease & 0.175  & 0.103  & 0.186  & 0.320  \\
		Chemical-Chemical & \textbf{0.483 } & 0.367 & \textbf{0.537 } & \textbf{0.713 } \\
		ALL   & 0.228  & 0.145  & 0.252  & 0.390  \\
		\bottomrule
	\end{tabular}%
	\label{tab10}%
\end{table}%

\begin{table}[htbp]
	\centering
	\caption{Experiments on the DRKG17k-21 subdataset}
	\begin{tabular}{lcccc}
		\toprule
		\textbf{Dataset} & \multicolumn{4}{c}{\textbf{DRKG17k-21}} \\
		\midrule
		\textbf{Task} & \multicolumn{4}{c}{\textbf{Link Prediction}} \\
		\midrule
		\textbf{Metric $\rightarrow$} & MRR   & H@1   & H@3   & H@10 \\
		\textbf{SubDataset $\downarrow$} &       &       &       &  \\
		\midrule
		Gene-Chemical & 0.333  & 0.258  & 0.363  & 0.480  \\
		Gene-Gene & 0.243  & 0.160  & 0.272  & 0.400  \\
		Gene-Disease & 0.222  & 0.138  & 0.223  & 0.447  \\
		Chemical-Chemical & \textbf{0.640 } & \textbf{0.549 } & \textbf{0.693 } & \textbf{0.811 } \\
		ALL   & 0.590  & 0.505  & 0.637  & 0.747  \\
		\bottomrule
	\end{tabular}%
	\label{tab11}%
\end{table}%

\begin{table}[htbp]
	\centering
	\caption{Results of baselines + NCL.}
	\begin{tabular}{lcccc}
		\toprule
		\textbf{Dataset} & \multicolumn{4}{c}{\textbf{PharmKG8k-28}} \\
		\midrule
		\textbf{Task} & \multicolumn{4}{c}{\textbf{Link Prediction}} \\
		\midrule
		\textbf{Metric $\rightarrow$} & MRR   & H@1   & H@3   & H@10 \\
		\textbf{Model  $\downarrow$} &       &       &       &  \\
		\midrule
		CompGCN & 0.193  & 0.110  & 0.216  & 0.352  \\
		SE-GNN & 0.206  & 0.120  & 0.232  & 0.374  \\
		CompGCN + NCL & \textbf{0.194 } & \textbf{0.112 } & \textbf{0.217 } & \textbf{0.354 } \\
		SE-GNN + NCL & \textbf{0.212 } & \textbf{0.127 } & \textbf{0.237 } & \textbf{0.377 } \\
		\bottomrule
	\end{tabular}%
	
	\label{tab7}%
\end{table}%

\begin{table}[htbp]
	\centering
	\caption{KGE with diffirent loss functions.}
	\begin{tabular}{lcccc}
		\toprule
		\textbf{Dataset} & \multicolumn{4}{c}{\textbf{PharmKG8k-28}} \\
		\midrule
		\textbf{Task} & \multicolumn{4}{c}{\textbf{Link Prediction}} \\
		\midrule
		\textbf{Metric $\rightarrow$} & MRR   & H@1   & H@3   & H@10 \\
		\textbf{Model  $\downarrow$} &       &       &       &  \\
		\midrule
		KGE + BCELoss & 0.222  & 0.141  & 0.244  & 0.379  \\
		KGE + MPLoss & 0.205  & 0.124  & 0.226  & 0.363  \\
		KGE + MRLoss & 0.163  & 0.082  & 0.176  & 0.326  \\
		\textbf{KGE+ NCLoss} & \textbf{0.228 } & \textbf{0.145 } & \textbf{0.252 } & \textbf{0.390 } \\
		\bottomrule
	\end{tabular}%
	
	\label{tab8}%
\end{table}%

\begin{figure*}[htbp]  
	\centering    
	
	\subfloat[Convergence of SE-GNN + NCL.] 
	{
		\begin{minipage}[t]{0.5\textwidth}
			\centering          
			\includegraphics[width=1\textwidth]{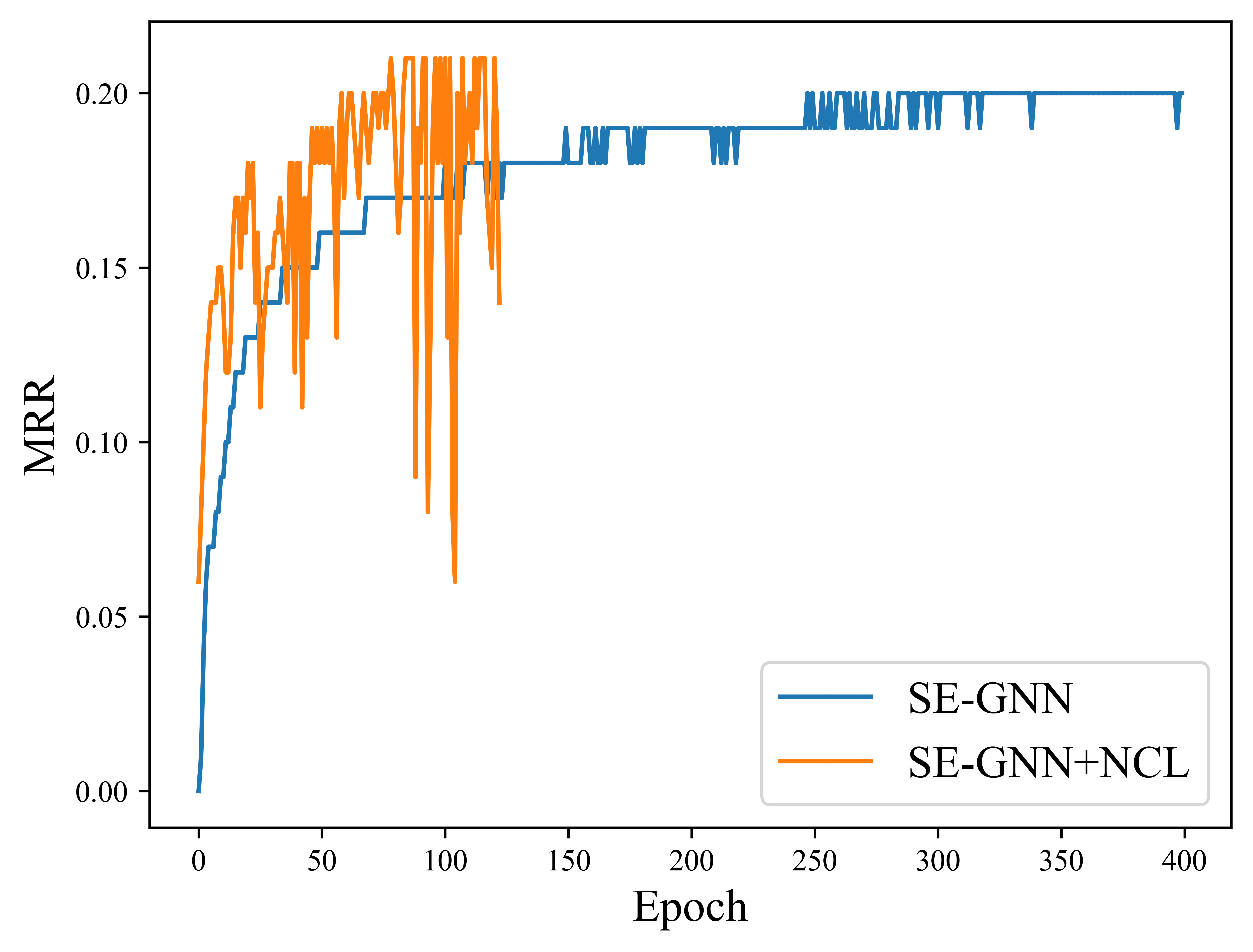}   
			\label{fig71}
		\end{minipage}%
	}
	\subfloat[Convergence of CompGCN + NCL.] 
	{
		\begin{minipage}[t]{0.5\textwidth}
			\centering          
			\includegraphics[width=1\textwidth]{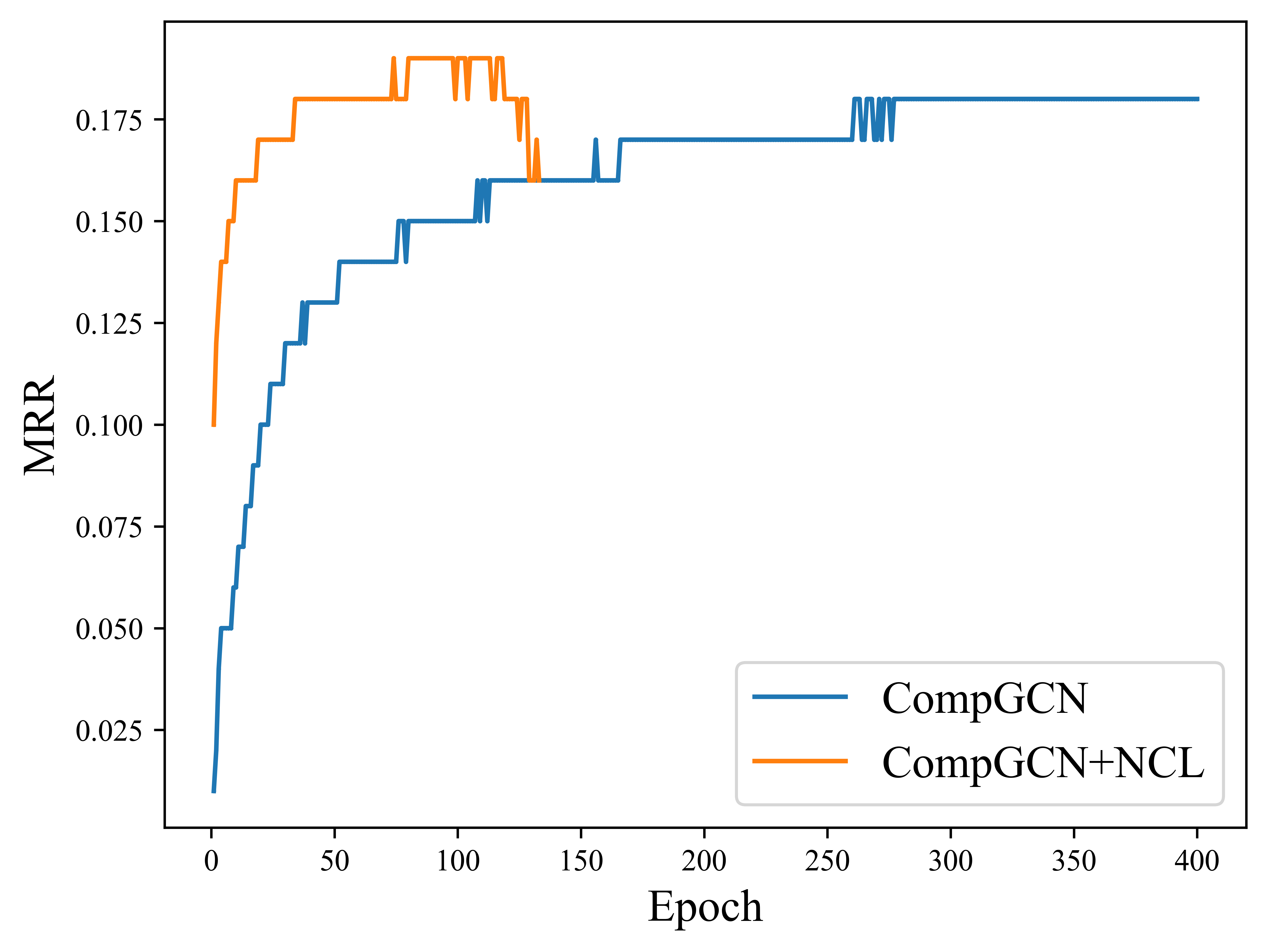}   
			\label{fig72}
		\end{minipage}%
	}
	\caption{Performance of the baseline models on PharmKG8k-28.} 
	\label{fig7}  
\end{figure*}

\subsection{Discussion}
\textbf{The Universality of NC-KGE.}
NC-KGE is a straightforward yet effective method that integrates contrastive learning with non-contrastive Knowledge Graph Embedding (KGE) techniques to enhance performance on biomedical knowledge graphs. In our study, we have integrated NC-KGE with existing models like CompGCN and SE-GNN, using PharmKG8k-28 as a representative example. The evaluation metrics for both CompGCN and SE-GNN demonstrated improvements, as outlined in Table \ref{tab7}. Furthermore, the evolution of Mean Reciprocal Rank (MRR) on the test set is depicted in Figure \ref{fig7}. NC-KGE not only enhances the accuracy of the models but also accelerates the convergence of CompGCN and SE-GNN. This suggests a general improvement resulting from node-based contrastive learning.

\textbf{The performance of NC-KGE with different similarity measure functions.}
In this paper, we consider the effect of different similarity measure functions on NC-KGE. As shown in Table \ref{tab6} of Appendix A, we consider 3 common classes of metric functions, including TransE based on translation, DistMult, SimplE and ComplEx based on tensor decomposition, ConvE with parameters based on convolutional networks. Among the five similarity measure functions, NC-KGE combined with ConvE has the best performance.

\textbf{The performance of NC-KGE under various temperature coefficients.}
In this section, we discuss the temperature hyperparameter for NC-KGE. From previous experience, smaller temperature coefficient usually makes the model tend to distinguish more difficult negative samples, and the learned embedding will be smoother. However, this may also lead to the model paying too much attention to difficult negative samples, resulting in insufficient attention to common negative samples, which is not conducive to learning good embeddings. As shown in Table \ref{tab6} of Appendix A, we show the temperature coefficient is set to 0.1, 0.2, 0.3, 0.5, 0.7, 0.8, 1.0, 1.1, 1.2, 1.3, 1.5 and dynamic adjustment. Among them, the temperature coefficient based on the dynamic adjustment strategy achieves better performance. However, when the temperature coefficient is set to a certain value, the performance of the model does not change significantly.

\textbf{The performance of NC-KGE with different number of negative samples.}
The number of negative samples is always an important discussion topic for self-supervised learning or unsupervised learning. In general, more negative samples will allow the model to see more diverse data points, which will improve performance. However, more negative samples also mean higher sampling cost. So we need to make a trade-off between the two. Here, we also show the performance of NC-KGE with different number of negative samples. As shown in Table \ref{tab6} of Appendix A, we find that NC-KGE achieves a decent performance when the number of negative samples for contrastive learning is 1000, and it performs best when the number is the whole.

\section{Conclusion}
In this study, we introduced a node-based contrastive knowledge graph embedding method called NC-KGE for relation prediction in biomedical knowledge graphs. NC-KGE is a simple and versatile approach that can be seamlessly integrated into existing KGE methods to enhance training convergence and improve overall performance. This distinctive feature allows NC-KGE to fully leverage advancements in state-of-the-art models.
In the future work, we will expand the application of NC-KGE to diverse downstream tasks, including but not limited to knowledge-guided molecular property prediction and the prediction of higher-order biomedical n-ary relationships. By leveraging semantically rich pre-trained node and relation embeddings, we aim to maximize the potential of NC-KGE in these domains.

\begin{acks}
	Thanks to the readers.
\end{acks}

\bibliographystyle{ACM-Reference-Format}
\bibliography{gnn_ref}

\appendix
\section{Appendix}
The detailed description of datasets is as follows:
\begin{itemize}
	\item \textbf{FB15k-237} \cite{toutanova2015observed} dataset contains knowledge base relation triples and textual mentions of Freebase node pairs, as used in the work published in \cite{toutanova2015observed}. The knowledge base triples are a subset of the FB15K \cite{bordes2013translating}, originally derived from Freebase. The inverse relations are removed in FB15k-237. To obtain node descriptions and types, we employ the datasets made available by \cite{daza2021inductive}. 

	\item \textbf{WN18RR} \cite{dettmers2018convolutional} is created from WN18 \cite{bordes2013translating}, which is a subset of WordNet. WN18 consists of 18 relations and 40,943 nodes. However, many text triples obtained by inverting triples from the training set. Thus WN18RR dataset \cite{dettmers2018convolutional} is created to ensure that the evaluation dataset does not have inverse relation test leakage. To obtain node descriptions and types, we employ the datasets made available by \cite{daza2021inductive}. 
	
	
	\item \textbf{PharmKG8k-28} is built based on PharmKG dataset \cite{10.1093/bib/bbaa344}. PharmKG is a multi-relational, attributed biomedical KG, composed of more than 500 000 individual interconnections between genes, drugs and diseases, with 29 relation types over a vocabulary of $\sim$ 8000 disambiguated nodes.  Each node in PharmKG is attached with heterogeneous, domain-specific information obtained from multi-omics data, i.e. gene expression, chemical structure and disease word embedding, while preserving the semantic and biomedical features. We obtained PharmKG8k-28 dataset after simple deduplication cleaning of PharmKG.

	\item \textbf{DRKG17k-21} is built based on Drug Repositioning Knowledge Graph (DRKG) \cite{drkg2020}. DRKG is a comprehensive biological knowledge graph relating genes, compounds, diseases, biological processes, side effects and symptoms.  DRKG includes information from six existing databases including DrugBank, Hetionet, GNBR, String, IntAct and DGIdb, and data collected from recent publications particularly related to Covid19.

	\item \textbf{BioKG72k-14} is built based on BioKG dataset \cite{DBLP:conf/cikm/WalshMN20}. BioKG is a new more standardised and reproducible biological knowledge graph which provides a compilation of curated relational data from open biological databases in a unified format with common, interlinked identifiers. BioKG can be used to train and assess the relational learning models in various tasks related to pathway and drug discovery. 
\end{itemize}

\begin{table*}[htbp]
	\centering
	\caption{Statistics of the datasets.}
	\begin{tabular}{lccccc}
		\toprule
		& \textbf{FB15k-237} & \textbf{WN18RR} & \textbf{PharmKG8k-28} & \textbf{DRKG17k-21} & \textbf{BioKG72k-14} \\
		Relations & 237   & 11    & 28    & 21    & 14 \\
		\midrule
		Nodes & 14,541 & 40,943 & 7,599 & 17,508 & 72,569 \\
		\midrule
		Triples & 310,116 & 93,003 & 480,500 & 587,425 & 1,157,739 \\
		\midrule
		& \multicolumn{5}{c}{Training} \\
		\cmidrule{2-6}    Nodes & 14,505 & 40,559 & 7,520 & 16,252 & 70,108 \\
		Triples & 272,115 & 86,835 & 384,880 & 470,526 & 927,348 \\
		& \multicolumn{5}{c}{Validation} \\
		\cmidrule{2-6}    Nodes & 9,809 & 5,173 & 6,637 & 8,186 & 43,806 \\
		Triples & 17,535 & 3,034 & 47,570 & 58,156 & 114,617 \\
		\midrule
		& \multicolumn{5}{c}{Test} \\
		\cmidrule{2-6}    Nodes & 10,348 & 5,323 & 6,628 & 8,242 & 43,989 \\
		Triples & 20,466 & 3,134 & 48,050 & 58,743 & 115,774 \\
		\bottomrule
	\end{tabular}%
	\label{tab2}%
\end{table*}%

\begin{table*}  
	\caption{Ablation experiments on the PharmKG8k-28 and FB15k-237 datasets. 'Neg100' means that the number of negative samples is 100. 'Tem0.1' means that the temperature coefficient is 0.1. 'DynamicTem' means to adopt the dynamic temperature coefficient strategy.}
	\centering
	\begin{tabular}{lcccc}
		
		\toprule
		\textbf{Dataset} & \multicolumn{4}{c}{\textbf{PharmKG8k-28}} \\
		\midrule
		\textbf{Task} & \multicolumn{4}{c}{\textbf{Link Prediction}} \\
		\midrule
		\textbf{Metric $\rightarrow$} & MRR   & H@1   & H@3   & H@10 \\
		\textbf{Model $\downarrow$} &       &       &       &  \\
		\midrule
		NC-KGE + TransE & 0.140  & 0.077  & 0.147  & 0.263  \\
		NC-KGE + DistMult & 0.169  & 0.097  & 0.177  & 0.317  \\
		NC-KGE + SimplE & 0.200  & 0.126  & 0.217  & 0.344  \\
		NC-KGE + ComplEx & 0.205  & 0.134  & 0.220  & 0.345  \\
		\textbf{NC-KGE +ConvE} & \textbf{0.228 } & \textbf{0.145 } & \textbf{0.252 } & \textbf{0.390 } \\
		\midrule
		NC-KGE + Neg100 & 0.214  & 0.126  & 0.241  & 0.385  \\
		NC-KGE + Neg500 & 0.214  & 0.126  & 0.241  & 0.384  \\
		NC-KGE + Neg1000 & 0.220  & 0.133  & 0.246  & 0.386  \\
		\textbf{NC-KGE +NegAll} & \textbf{0.228 } & \textbf{0.145 } & \textbf{0.252 } & \textbf{0.390 } \\
		\midrule
		NC-KGE + Tem0.1 & 0.218  & 0.137  & 0.239  & 0.376  \\
		NC-KGE + Tem0.2 & 0.220  & 0.138  & 0.242  & 0.379  \\
		NC-KGE + Tem0.3 & 0.217  & 0.135  & 0.240  & 0.377  \\
		NC-KGE + Tem0.5 & 0.220  & 0.138  & 0.242  & 0.379  \\
		NC-KGE + Tem0.7 & 0.225  & 0.142  & 0.247  & 0.386  \\
		NC-KGE + Tem0.8 & 0.223  & 0.140  & 0.247  & 0.383  \\
		NC-KGE + Tem1.0 & 0.225  & 0.142  & 0.248  & 0.386  \\
		NC-KGE + Tem1.1 & 0.225  & 0.143  & 0.248  & 0.386  \\
		NC-KGE + Tem1.2 & 0.227  & 0.145  & 0.250  & 0.384  \\
		NC-KGE + Tem1.3 & 0.225  & 0.142  & 0.250  & 0.386  \\
		NC-KGE + Tem1.5 & 0.225  & 0.142  & 0.249  & 0.386  \\
		\textbf{NC-KGE +DynamicTem} & \textbf{0.228 } & \textbf{0.145 } & \textbf{0.252 } & \textbf{0.390 } \\
		\bottomrule
	\end{tabular}%
	\centering
	\begin{tabular}{lcccc}
		\toprule
		\textbf{Dataset} & \multicolumn{4}{c}{\textbf{FB15k-237}} \\
		\midrule
		\textbf{Task} & \multicolumn{4}{c}{\textbf{Link Prediction}} \\
		\midrule
		\textbf{Metric $\rightarrow$} & MRR   & H@1   & H@3   & H@10 \\
		\textbf{Model $\downarrow$} &       &       &       &  \\
		\midrule
		NC-KGE + TransE & 0.321  & 0.235  & 0.352  & 0.494  \\
		NC-KGE + DistMult & 0.322  & 0.235  & 0.353  & 0.499  \\
		NC-KGE + SimplE & 0.318  & 0.232  & 0.347  & 0.494  \\
		NC-KGE + ComplEx & 0.323  & 0.236  & 0.354  & 0.500  \\
		\textbf{NC-KGE +ConvE} & \textbf{0.362 } & \textbf{0.270 } & \textbf{0.392 } & \textbf{0.542 } \\
		\midrule
		NC-KGE + Neg100 & 0.342  & 0.248  & 0.377  & 0.531  \\
		NC-KGE + Neg500 & 0.352  & 0.259  & 0.387  & 0.538  \\
		NC-KGE + Neg1000  & 0.354  & 0.258  & 0.388  & 0.539  \\
		\textbf{NC-KGE +NegAll} & \textbf{0.362 } & \textbf{0.270 } & \textbf{0.392 } & \textbf{0.542 } \\
		\midrule
		NC-KGE + Tem0.1 & 0.347  & 0.255  & 0.382  & 0.531  \\
		NC-KGE + Tem0.2 & 0.348  & 0.255  & 0.384  & 0.530  \\
		NC-KGE + Tem0.3 & 0.348  & 0.258  & 0.381  & 0.528  \\
		NC-KGE + Tem0.5 & 0.348  & 0.257  & 0.381  & 0.529  \\
		NC-KGE + Tem0.7 & 0.348  & 0.254  & 0.383  & 0.534  \\
		NC-KGE + Tem0.8 & 0.347  & 0.256  & 0.381  & 0.531  \\
		NC-KGE + Tem1.0 & 0.349  & 0.257  & 0.382  & 0.535  \\
		NC-KGE + Tem1.1 & 0.349  & 0.256  & 0.383  & 0.536  \\
		NC-KGE + Tem1.2 & 0.348  & 0.258  & 0.384  & 0.537  \\
		NC-KGE + Tem1.3 & 0.348  & 0.259  & 0.385  & 0.536  \\
		NC-KGE + Tem1.5 & 0.349  & 0.258  & 0.386  & 0.536  \\
		\textbf{NC-KGE +DynamicTem} & \textbf{0.362 } & \textbf{0.270 } & \textbf{0.392 } & \textbf{0.542 } \\
		\bottomrule
	\end{tabular}%
	\label{tab6}
\end{table*}

\begin{figure*}[htbp]  
	\centering    
	
	\subfloat[Convergence on FB15k-237.] 
	{
		\begin{minipage}[t]{0.5\textwidth}
			\centering          
			\includegraphics[width=1\textwidth]{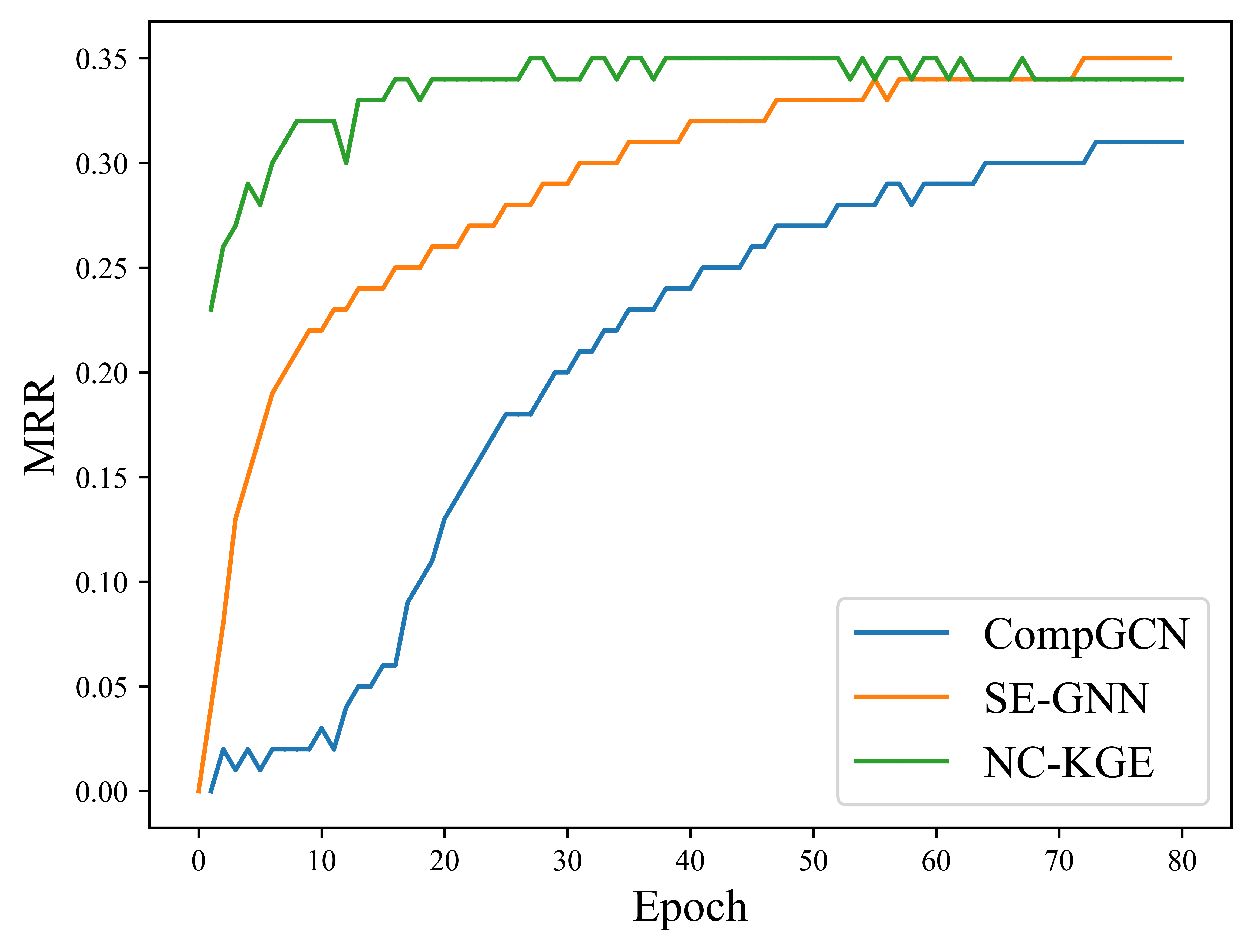}   
			\label{fig81}
		\end{minipage}%
	}
	\subfloat[Convergence on WN18RR.] 
	{
		\begin{minipage}[t]{0.5\textwidth}
			\centering          
			\includegraphics[width=1\textwidth]{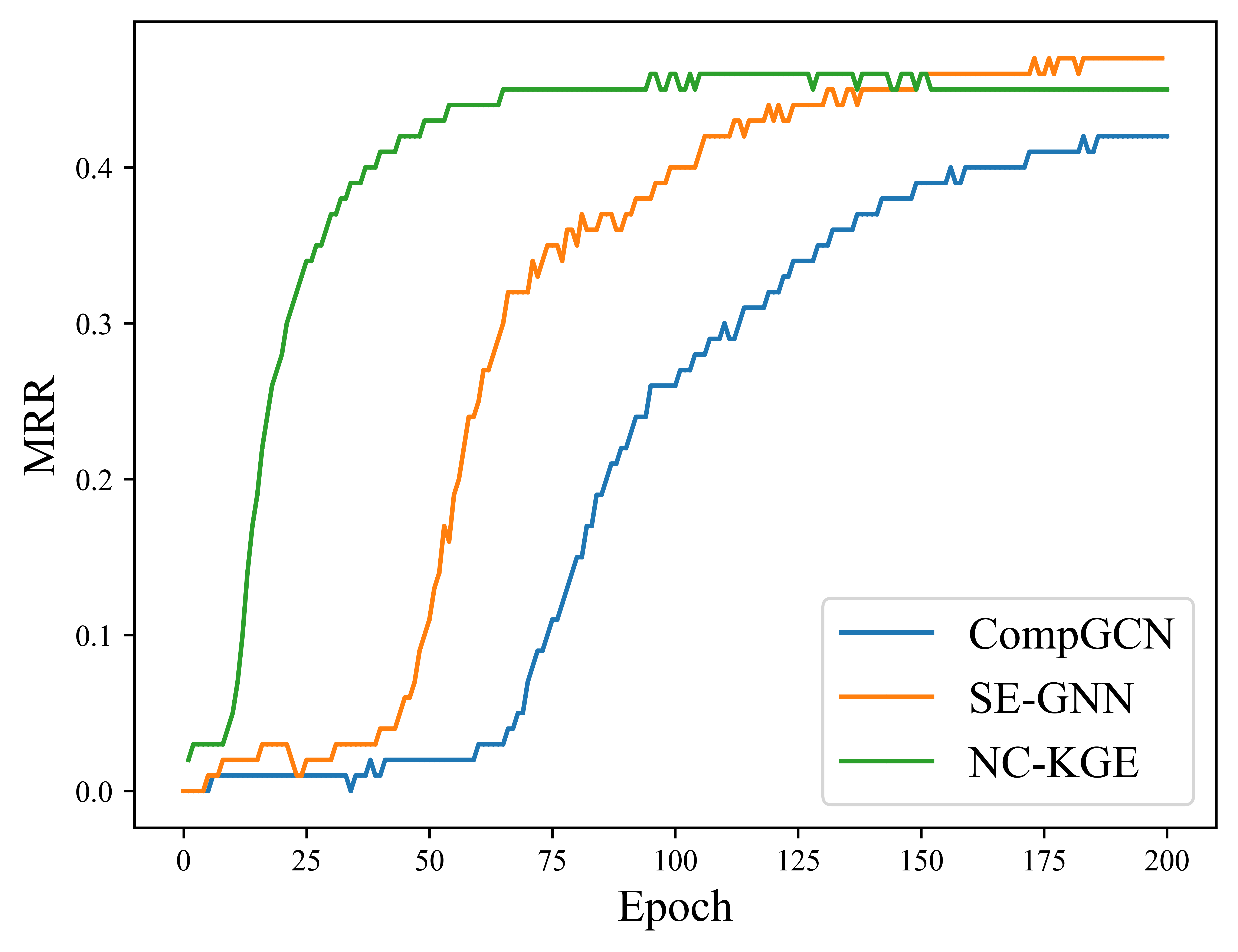}   
			\label{fig82}
		\end{minipage}%
	}
	\caption{Performance of the models on FB15k-237 and WN18RR.} 
	\label{fig8}  
\end{figure*}

\end{document}